\documentclass[
 aip,
 amsmath,amssymb,
 reprint,%
]{revtex4-1}

\usepackage{graphicx}% Include figure files
\usepackage{dcolumn}% Align table columns on decimal point
\usepackage{bm}% bold math
\usepackage[utf8]{inputenc}
\usepackage[T1]{fontenc}
\usepackage{mathptmx}
\usepackage{etoolbox}

\usepackage{textcomp}%
\usepackage{booktabs}%
\usepackage{algorithm}%
\usepackage{algorithmicx}%
\usepackage{algpseudocode}%
\usepackage{listings}%
\usepackage{tikz}
\usepackage{colortbl}
\usepackage{tabularray}
\usepackage{siunitx}
\usepackage{multirow} % for multirow cells in tables
\usepackage{array}    % for centering text in table columns
\makeatletter
\def\@email#1#2{%
 \endgroup
 \patchcmd{\titleblock@produce}
  {\frontmatter@RRAPformat}
  {\frontmatter@RRAPformat{\produce@RRAP{*#1\href{mailto:#2}{#2}}}\frontmatter@RRAPformat}
  {}{}
}%
\makeatother

\begin{document}

\preprint{AIP/123-QED}

\title{A Hybrid Neural Architecture:\\ Online Attosecond X-ray Characterization}% Force line breaks with \\
\author{J. Hirschman}
\email{jhirschm@stanford.edu}
\affiliation{Department of Applied Physics, Stanford University, 348 Via Pueblo, Stanford, CA 94305, USA}
\affiliation{SLAC National Accelerator Laboratory, 2575 Sand Hill Rd, Menlo Park, CA 94025, USA}
\author{B. Mencer}%
\affiliation{SLAC National Accelerator Laboratory, 2575 Sand Hill Rd, Menlo Park, CA 94025, USA}
\affiliation{Department of Computer Science and Engineering, University of California, Santa Cruz, 1156 High St, Santa Cruz, CA 95064, USA}

\author{A. Shackelford}
\affiliation{SLAC National Accelerator Laboratory, 2575 Sand Hill Rd, Menlo Park, CA 94025, USA}

\author{R. Obaid}
\affiliation{SLAC National Accelerator Laboratory, 2575 Sand Hill Rd, Menlo Park, CA 94025, USA}

\author{R. Coffee}
\affiliation{SLAC National Accelerator Laboratory, 2575 Sand Hill Rd, Menlo Park, CA 94025, USA}

\date{\today}% It is always \today, today,

\begin{abstract}
The emergence of high-repetition-rate x-ray free-electron lasers, such as SLAC’s LCLS-II, serve as our canonical example for autonomous controls that necessitate high-throughput diagnostics paired with streaming computational pipelines capable of single-shot analysis with extremely low latency. 
We present the Deterministic Characterization with an Integrated Parallelizable Hybrid Resolver architecture, a hybrid machine learning framework designed for fast, accurate analysis of XFEL diagnostics using angular streaking-based sinogram images. 
This architecture integrates convolutional neural networks and bidirectional long short-term memory models to denoise input, identify x-ray sub-spike features, and extract sub-spike relative delays with sub-30 attosecond temporal resolution. 
Deployed on low-latency hardware, it achieves over 10~kHz throughput with \SI{168.3}{\micro\second} inference latency, indicating scalability to 14~kHz with FPGA integration.
By transforming regression tasks into classification problems and leveraging optimized error encoding, we achieve high precision with low-latency performance that is critical for real-time streaming event selection and experimental control feedback signals. 
This represents a key development in real-time control pipelines for next-generation autonomous science, generally, and high repetition-rate x-ray experiments in particular.

\end{abstract}

\maketitle

\section{Introduction}
\label{sec:intro}
Real-time data processing for rapid, online decision-making is driving a paradigm shift in applied computing, one that moves away from reliance on large cloud computing infrastructures and moves toward so called Edge-Machine Learning (Edge-ML).
This evolution emphasizes the need for local, high-throughput, low-latency processing pipelines.
Beyond algorithmic innovation, Edge-ML demands compute hardware optimized for specific applications, requiring a co-design approach to take optimal advantage of hardware heterogeneity~\cite{emani2022comprehensive, osti_1604756}.
The U.S. Department of Energy (DOE) has identified this need in its report on Extreme Heterogeneity, stressing the importance of algorithms and hardware co-optimized to meet application-specific demands~\cite{osti_1473756} as well as integrated compute architectures for ML in its Artificial Intelligence for Science report~\cite{osti_1604756}.
Edge-ML and heterogeneous computing have already gained traction across diverse applications, from autonomous vehicles and defense systems to tokamak fusion plasma stabilization, future Higgs factories, and advanced light sources like x-ray free-electron lasers (XFELs)~\cite{ibn2022next, churchill2020machine, churchill2023accelerating, ruvinsky2021accelerating,o2023automatic, shousha2025unified, rothstein2024initial, apresyan2023detector, gonski2024embedded, yue2025variational,  therrien2019machine, cryan2022development}.
These cutting-edge fields share a common need: precise, high-throughput, and low-latency data processing.
Furthermore, their data processing pipelines often require denoising, feature identification, and feature extraction.
Here, we present a novel multi-component ML framework--Deterministic Characterization with an Integrated Parallelizable Hybrid Resolver (DCIPHR)--specifically designed for high-repetition-rate data acquisition, diagnostics and real-time control pipelines.

To demonstrate DCIPHR's inference capabilities, we target x-ray production at SLAC National Accelerator Laboratory’s Linac Coherent Light Source (LCLS), the world’s first hard x-ray free-electron laser and a leading facility for ultrafast x-ray science.
LCLS and its upgraded version, LCLS-II, enable groundbreaking research across atomic, molecular, and optical (AMO) physics, ultrafast physics, crystallography, biology, chemistry, and more~\cite{young2010femtosecond, OstromChemicalBonds, BoutetCrystallography, boutet2018x, stankus2019ultrafast}.
The ongoing LCLS-II upgrades are unlocking finer spatial resolution through an expanded x-ray photon energy range, delivering attosecond x-ray pulses for probing inner-core electron dynamics with the ability to measure electron-electron coherences, and scaling repetition rate in stages from 120 Hz to 1 MHz, already meeting major milestones~\cite{duris2020tunable, cryan2022development}.
These advances demand diagnostics coupled with processing that deliver attosecond precision, broad spectral adaptability, and high throughput--all with realtime constraints.
The adapted DCIPHR model paired with LCLS-II's recently commissioned single-shot x-ray detectors meet these challenges by integrating convolution neural networks (CNNs)~\cite{lecun1990proc, lecun2015deep}, fully connected layers, and bidirectional long short-term memory (BiLSTM) models~\cite{hochreiter1997long, gers2002learning, graves2005framewise} to improve detector signal-to-noise, rapidly identify x-ray substructure features with accuracy greater than 96\%, and enable sub-30~attosecond x-ray substructure temporal resolution when trained on synthetic data.
Moreover, DCIPHR can achieve inter-model pipelined throughput on Groq AI Inference cards of 14 kHz, with the modular architecture enabling further speed-up when deployed heterogeneously, making it well-suited for integration into truly heterogeneously composed real-time data processing engines.

\section{Diagnostic \& Methods}
\label{sec:background}
\subsection{X-Rays \& Diagnostics}
\begin{figure*}[ht!]
\centering\includegraphics[width=1\textwidth]{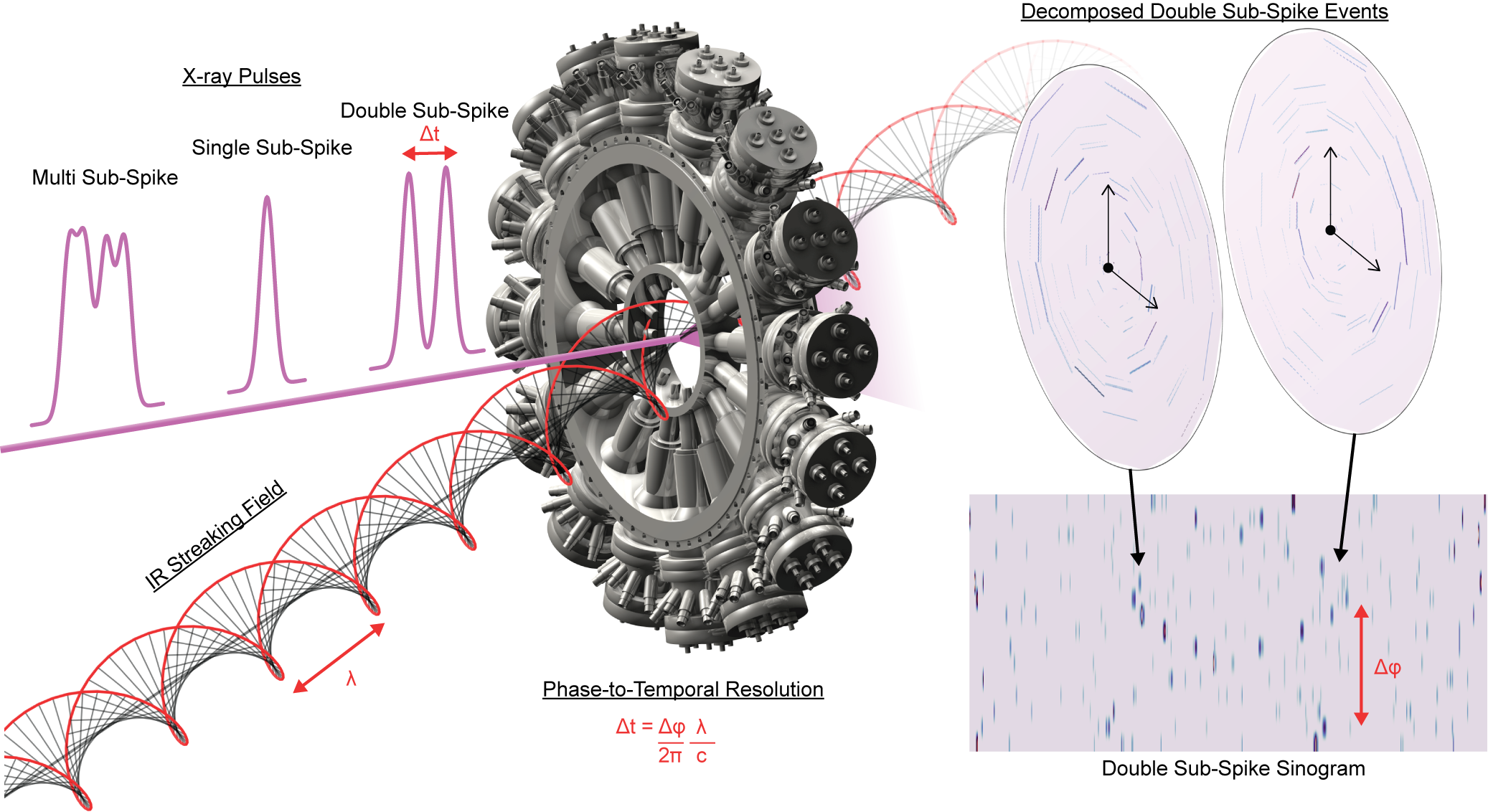}
\caption{\textbf{X-ray Detection and Aggregated Readout}
Circularly polarized IR streaking field creates a momentum shift in the x-ray photonionized electrons at the interaction point in the MRCO detector. At the front of the pulse train is a 2 SASE sub-spike x-ray shot, resulting in two ionization events that produce a sinogram with 2 sinusoids separated vertically, an effective phase shape which can be mapped to temporal separation of the sub-spikes based on streaking laser wavelength.}
\label{fig:main}
\end{figure*}

XFEL facilities traditionally generate x-rays via self-amplification of spontaneous emission (SASE), resulting in few femtosecond down to attosecond-scale x-ray substructures called SASE sub-spikes that collectively form the temporal profile of each pulse (see Fig.~\ref{fig:main})~\cite{cryan2022development,hartmann2018attosecond,marinelli2017experimental,huang2017generating, malyzhenkov2020single, lutman2018high}.
Due to the stochastic nature of SASE, the number, distribution, and photon energy of these sub-spikes fluctuate on a shot-to-shot basis~\cite{lutman2013experimental}.
In an alternative mode of operation called x-ray laser enhanced attosecond pulse generation (XLEAP), isolated attosecond x-ray pulses are produced using enhanced SASE (ESASE)~\cite{zholents2005method, duris2020tunable}.
These pulses still have variation in the longitudinal pulse profile from shot-to-shot~\cite{duris2020tunable}.
For both modes, the x-ray diagnostics must elucidate the underlying substructure in the x-ray pulse, necessitating a single-shot diagnostic.

To meet the x-ray pulse reconstruction challenges, the Multi-Resolution COokiebox (MRCO) detector was developed~\cite{walter2021multi}.
This system performs simultaneous energy and angle-resolved electron time-of-flight spectroscopy using a circular array of spectrometers (see Fig.~\ref{fig:main}), each recording the arrival times of electrons photoionized from gas phase samples at the interaction point~\cite{walter2021multi, hartmann2018attosecond}.
Via attosecond angular streaking, the temporal intensity profile of the x-ray pulses can be mapped onto the angular emission distribution of the ionized electrons (see Fig.~\ref{fig:main})
by concurrently introducing a circularly polarized infrared (IR) laser pulse at the interaction region, called ``optical dressing'' of the atoms~\cite{eckle2008attosecond, itatani2002attosecond, hartmann2018attosecond, kazansky2016interference, constant1997methods}.
These signals are converted to energy and aggregated into sinogram images of 16x512--16 detector channels by 512 energy bins.
Fig.~\ref{fig:main} illustrates this process with an example of a two sub-spike x-ray photoionization and the associated time-shifted kicks which ultimately produce a sinogram with vertically shifted sinusoids.
For this two sub-spike case, the vertical phase difference ($\Delta\phi$) of the sinusoids maps to the relative sub-spike arrival time ($\Delta t$) based on the time-variation of the dressing laser carrier field as shown in Eq.~\ref{eq:temporal} and Fig.~\ref{fig:main}

\begin{equation}
\label{eq:temporal}
    \Delta t = \frac{\Delta \phi T }{2\pi}
\end{equation}

where \( T = \frac{\lambda}{c} \) is the laser period, determined by the wavelength \( \lambda \) and the speed of light \( c \).  
This two sub-spike example marks a particularly interesting use-case for x-ray pump/x-ray probe where x-ray pulses are both the excitation and the probe mechanisms~\cite{hartmann2018attosecond, guo2024experimental, robles2025spectrotemporal}.

The MRCO detector, along with other high-speed detectors stationed at SLAC's LCLS-II, is projected to generate terabits of data per second as LCLS-II's operation rate approaches 1~MHz. 
This creates immense pressure on the data processing pipeline and storage facilities~\cite{thayer2017building}.
Online processing of the x-ray shots will be required for fast veto or, more interestingly, for real-time data sorting that could enabling new experimental methods that leverage the stochasticity of SASE. 
For example, traditionally scanned x-ray pump/x-ray probe experiments require fine delay control; however, single shot diagnostics would enable a free-running SASE mode to leverage the inherently noise-driven fluctuations in the x-ray temporal structure to stochastically produce the data, naturally averaging over aliasing effects and spurious correlations~\cite{lutman2013experimental, lutman2016fresh, marinelli2015high}.

Real-time characterization, sorting, and reconstruction of high-rate x-ray shots from MRCO require algorithms that are capable of extracting essential features from complex sinogram images, where the number of sinusoids is not predetermined.
The sinograms are created only after the multiple channels of the MRCO detector have been brought together; we note that each of the channels is transferred to the data center over individual optical fiber, each landing on different nodes in the data reduction pipeline system.
As such we must first do the so called ``event building'' that associates the independent channel signals for a given shot together as an aggregated multi-channel signal.
Post-aggregation processing involves denoising the resulting sinograms, identifying the number of sinusoids in an x-ray shot, and extracting the SASE sub-spike characteristics, where relevant.
Hartmann \textit{et al.}~successfully identified the number of sinusoids in the sinogram and reconstructed two sub-pulse cases using an iterative convolution kernel method~\cite{hartmann2018attosecond}; however, this method does not support high-throughput and low-latency processing.
Other approaches have explored standard image processing techniques and machine learning (ML) methods, including neural networks (NN)~\cite{kumar2020purifying, dingel2022artificial, funke2024capturing, heider2019megahertz, li2018characterizing}. 
Yet, a scalable and latency-aware ML pipeline capable of real-time x-ray substructure identification and reconstruction remains elusive, until now.

% \begin{table*}[]
% \begin{centering}
% \caption{DCIPHR module summary with architecture, purpose, and primary evaluation metric}
% \label{tab:DCIPHRComponentSummary}
% \begin{tabular}{|l|l|l|l|}
% \hline
% \rowcolor[HTML]{C0C0C0} 
% \textbf{Module} & \textbf{Purpose} & \textbf{Architecture} & \textbf{Metric} \\ \hline

% \textbf{Denoiser (1)} & 
% Suppress noise & 
% \begin{tabular}[c]{@{}l@{}}CNN + \\ Autoencoder \end{tabular} &
% \begin{tabular}[c]{@{}l@{}}SSIM-G\end{tabular} \\ \hline

% \textbf{Classifier (2)} & 
% \begin{tabular}[c]{@{}l@{}}Count sub-spikes \\ (sinusoids) in sinogram\end{tabular} &
% \begin{tabular}[c]{@{}l@{}}BiLSTM + \\ FC Layers\end{tabular} &
% \begin{tabular}[c]{@{}l@{}}Cross-Entropy \\ Loss\end{tabular} \\ \hline

% \begin{tabular}[c]{@{}l@{}}\textbf{Single-Pulse} \\ \textbf{Regressor (3)}\end{tabular} &
% Predict $\phi$ for single pulse & 
% ResNet-18 & 
% \begin{tabular}[c]{@{}l@{}}Error encoding \\ + RMSE\end{tabular} \\ \hline

% \begin{tabular}[c]{@{}l@{}}\textbf{Double-Pulse} \\ \textbf{Regressor (4)}\end{tabular} & 
% Predict $\Delta\phi$ from two pulses & 
% ResNet-34 & 
% \begin{tabular}[c]{@{}l@{}}Error encoding \\ + RMSE\end{tabular} \\ \hline

% \end{tabular}
% \end{centering}
% \end{table*}

\begin{table*}
\caption{DCIPHR module summary with architecture, purpose, and primary evaluation metric}
\label{tab:DCIPHRComponentSummary}
\begin{ruledtabular}
\begin{tabular}{llll}
\textbf{Module} & \textbf{Purpose} & \textbf{Architecture} & \textbf{Metric} \\ \hline
Denoiser (1) &
Suppress noise &
CNN + Autoencoder &
SSIM-G \\

Classifier (2) &
Count sub-spikes (sinusoids) in sinogram &
BiLSTM + FC Layers &
Cross-Entropy Loss \\

Single-Pulse Regressor (3) &
Predict $\phi$ for single pulse &
ResNet-18 &
Error encoding + RMSE \\

Double-Pulse Regressor (4) &
Predict $\Delta\phi$ from two pulses &
ResNet-34 &
Error encoding + RMSE \\
\end{tabular}
\end{ruledtabular}
\end{table*}

\subsection{Models \& Metrics}
\begin{figure*}[ht!]
\centering\includegraphics[width=1\textwidth]{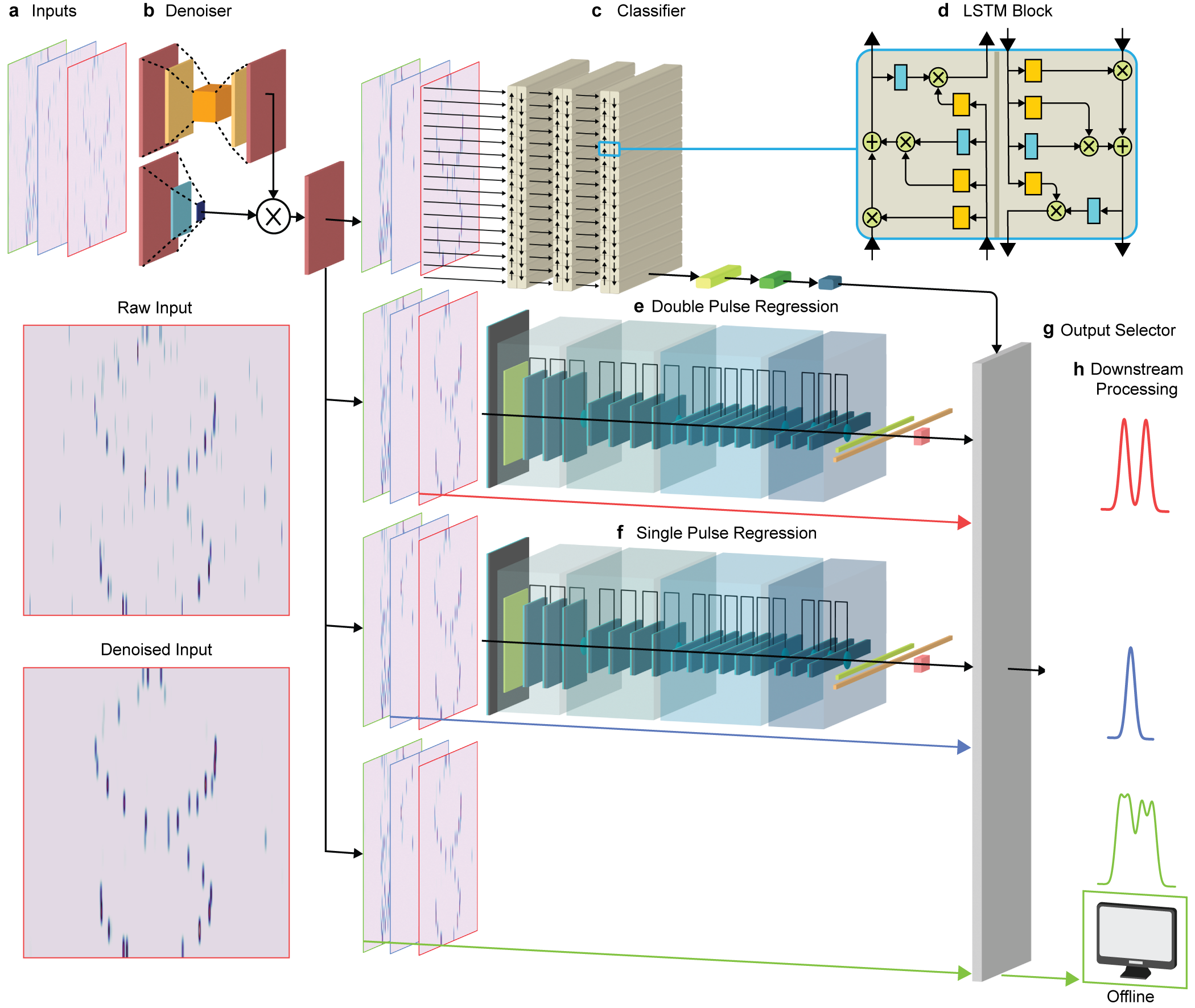}
\caption{\textbf{DCIPHR Architecture}
a) noisy input sinogram images are fed into b) the autoencoder-based denoiser; these denoised sinograms are then concurrently streamed to the c) SASE sub-spike classifier network, the e) double sub-spike regression network, and the f) single sub-spike regression network where d) shows the details of the LSTM block inside the classifier network; the output from the classifier is used as the select input for the g) output selector which routes both the output from the classifier as well as the properly selected double sub-spike or single sub-spike output for downstream analysis.}
\label{fig:model}
\end{figure*}

DCIPHR is a modular and expandable pipeline architecture targeted at deterministic classification and extraction of intertwined features at high resolution.
This adapted DCIPHR network for MRCO is trained on synthetic data from CookieSimSlim~\cite{cookiesimslim, gouin2022data}, which generates datasets that emulate attosecond angular streaking experiments performed with the MRCO (see Supplementary Material).
These simulations produce noisy sinogram images, ground truth probability distributions, and associated sinusoid parameters such as number of sinusoids or sub-spikes and the corresponding vertical phases of these spikes.
Using this data, DCIPHR operates in several steps as shown in Fig.~\ref{fig:model}: 1) inputs are mapped from noisy sinogram to the probability distribution; the denoised sinograms are fed concurrently into 2) the BiLSTM-based SASE sub-spike classifier to determine if there are 0, 1, 2, 3, or 4+ sub-spikes and into 3) and 4) ResNet-based regression networks for phase information retrieval for the single and double sub-spike cases respectively.  
The BiLSTM SASE sub-spike classifier then both serves as the control input for the output selector and is itself an output for downstream analysis. 
A block diagram is shown in Fig.~S1 in the Supplementary , and a summary of the models and metrics is shown in Table~\ref{tab:DCIPHRComponentSummary}.

\subsubsection{Denoiser}
The denoiser module (Fig.~\ref{fig:model}b) reconstructs a cleaned probability distribution from the noisy and incomplete sinogram input (Fig.~\ref{fig:model}a) to suppress noise and enhance the sinusoidal features that are relevant to sub-spike analysis.  
This model consists of an encoder-decoder architecture trained as a denoising autoencoder, where the input is a noisy sinogram  (\(X_\texttt{img}\)) and the target output is the corresponding clean distribution (\(Y_\texttt{pdf}\)), in conjunction with a classifier--known here as the zero-pulse detector--that handles the 0 sub-spike edge case. 

The encoder extracts hierarchical features through a sequence of convolutional layers, compressing the input into a latent representation that captures the underlying sinusoidal structures while suppressing noise.
The decoder reconstructs the denoised distribution from this latent space using transposed convolutions.

The input is a single-channel sinogram of dimension \(16 \times 512\).
The encoder consists of three convolutional layers:
\begin{itemize}
    \item A 2D convolution with 16 filters of size \(3 \times 3\), padding 2, followed by ReLU activation.
    \item A 2D convolution with 32 filters of size \(3 \times 3\), padding 1, followed by ReLU.
    \item A 2D convolution with 64 filters of size \(3 \times 3\), padding 1, followed by ReLU.
\end{itemize}

The decoder uses three transposed convolutional layers:
\begin{itemize}
    \item A transposed convolution with 32 filters of size \(3 \times 3\), padding 1, followed by ReLU.
    \item A transposed convolution with 16 filters of size \(3 \times 3\), padding 1, followed by ReLU.
    \item A final transposed convolution with a single output channel, kernel size \(3 \times 3\), padding 2, followed by a sigmoid activation.
\end{itemize}

To suppress false reconstructions, the zero-pulse detector determines whether any sinusoidal features are present in the input. 
Its architecture includes:
\begin{itemize}
    \item A 2D convolutional layer with 16 filters (\(3 \times 3\), stride 1, padding 1), followed by ReLU.
    \item A max-pooling layer (\(3 \times 3\), stride 2).
    \item A second convolutional layer with 32 filters (\(3 \times 3\), stride 1, padding 1), followed by ReLU.
    \item A max-pooling layer (\(2 \times 2\), stride 2).
    \item A fully connected layer reducing to 4 units, followed by ReLU.
    \item A final fully connected layer outputting a scalar value for binary classification.
\end{itemize}

If the detector identifies an image as having no sinusoids, the corresponding autoencoder output is zeroed out.

While the classifier is evaluated with binary cross-entropy, the autoencoder portion adopts the Structural Similarity Index Measure (SSIM), a metric commonly used in medical imaging, as in positron emission tomography (PET), to compare image fidelity~\cite{wang2004image,amirrashedi2021deep, mudeng2022prospects}. 
With access to noisy input ($X_{\texttt{img}}$), denoised output ($\hat Y_{\texttt{pdf}}$), and ground truth ($ Y_{\texttt{pdf}}$) images, we further define the SSIM Gain (SSIM-G) metric as the percentage improvement in SSIM between the denoised and ground truth images relative to the noisy input and ground truth. 
The SSIM-G metric effectively captures the denoiser’s ability to enhance image structure beyond noise suppression alone.

\begin{equation}
\begin{aligned}
    \texttt{SSIM-G}  = 
    \frac{
        \texttt{SSIM}(\hat{Y}_{\text{pred}}, Y_{\text{pdf}}) 
        - \texttt{SSIM}(X_{\text{img}}, Y_{\text{pdf}})
    }{
        \texttt{SSIM}(X_{\text{img}}, Y_{\text{pdf}})
    }
\end{aligned}
\end{equation}

\subsubsection{Sub-spike Classifier}
The classification component of the DCIPHR network (Fig.~\ref{fig:model}c) operates on the denoised sinogram outputs to determine the number of sub-spikes present.  
Building on the success of LSTM models for feature extraction from 2D and 3D images~\cite{liu2017bidirectional, Byeon_2015_CVPR, NEURIPS2022_f9d7d6c6}, we use a BiLSTM network to scan vertically across the sinogram channels, capturing inter-channel dependencies that reveal complex sinusoidal patterns.  
This scanning both ``up'' and ``down'' the image in the vertical axis is a distinctive adaptation that enables detection of spatial correlations across adjacent detector channels.  
Unlike traditional BiLSTM models designed for temporal sequences, this configuration interprets spatial structures within 2D sinograms, allowing it to capture overlapping sinusoids, phase shifts, and other spatial variations across the energy axis.  
By leveraging vertical inter-channel relationships, the BiLSTM enhances the model's ability to differentiate between subtle changes in sinusoid positioning and accurately determine the number of sub-spikes.  

The BiLSTM consists of three stacked layers, each with a hidden size of 128.
The LSTM output is passed through two fully connected layers with 32 and 64 units respectively, using ReLU activations.
Dropout layers with a rate of 20\% are applied during training to improve generalization.
Final classification is performed using a softmax-activated output layer.

Training uses Cross Entropy Loss with the Adam optimizer (initial learning rate: 0.0001). 
A learning rate scheduler reduces the rate on plateau to promote stable convergence.

\subsubsection{Phase Regression}
To address the one-pulse and two-pulse phase regression tasks, recalling that phase here is synonymous with time or delay, DCIPHR employs ResNet architectures, chosen for their robust representational capacity and performance in both classification and regression tasks~\cite{He_2016_CVPR, torchvision2016}.
ResNet's residual connections mitigate vanishing gradient issues~\cite{he2016deep}, making them particularly well-suited for capturing subtle, hierarchical features, such as sinusoidal phase differences in high-dimensional sinograms.

Specifically, DCIPHR employs ResNet-18 and ResNet-34 architectures based on the Torch Vision implementation~\cite{torchvision2016} to perform phase regression on denoised sinograms for the single sub-spike regressor--predicting the vertical phase of an individual sinusoid--and the double sub-spike regressor--estimating the phase difference between two sinusoids--, respectively.

ResNet-18 consists of an initial $7 \times 7$ convolutional layer with 64 filters, stride 2, followed by batch normalization, ReLU activation, and $3 \times 3$ max pooling.  
This is followed by four residual stages, each containing two Basic Blocks with increasing channel widths: 64, 128, 256, and 512.  
A global average pooling layer reduces the spatial dimensions, followed by a fully connected layer to output 2000 phase classes.  

ResNet-34 shares the same initial layers as ResNet-18 but expands the depth of each residual stage:  
three Basic Blocks at 64 channels, four at 128, six at 256, and three at 512.  
After global average pooling, the final output layer maps to 4000 phase difference classes.

Each Basic Block comprises two $3 \times 3$ convolutional layers with batch normalization and ReLU activations.  
Residual connections directly add the block’s input to its output.  
When spatial dimensions or channel counts differ between layers, a $1 \times 1$ convolution is applied to the shortcut path for alignment.

Here, we focus on the more complex double sub-spike case and reserve the single sub-spike case for the Supplementary Material.
Binary cross-entropy loss with logits optimizes model performance, while the true phase differences are mapped to the 4000-class discretized range using 
\begin{equation}
    \label{eq:phase_dif}
    \mathbf{y} = \text{one-hot}\left(\frac{\arccos(\cos(\Delta\phi))}{\pi}\right)
\end{equation}
where the final phase error is calculated using root mean-squared error (RMSE) and mapped to temporal separation via Eq.~\ref{eq:temporal}.

\section{Results}
\label{sec:results}
Each of DCIPHR's modules demonstrate success in their respective tasks with the denoiser on average showing a 29\% improvement in signal quality, the sub-spike classifier achieving greater than 96\% accuracy for each class, and the double sub-spike phase difference regressor achieving a phase difference error of 0.04652
radians, which corresponds to 29.6 attoseconds when calculated for a \SI{1.2}{\micro\meter} streaking laser (Eq.~\ref{eq:temporal}). 

\subsection{Denoiser}
\label{sec:denoiser}

\begin{figure}[ht!]
\centering\includegraphics[width=.5\textwidth]{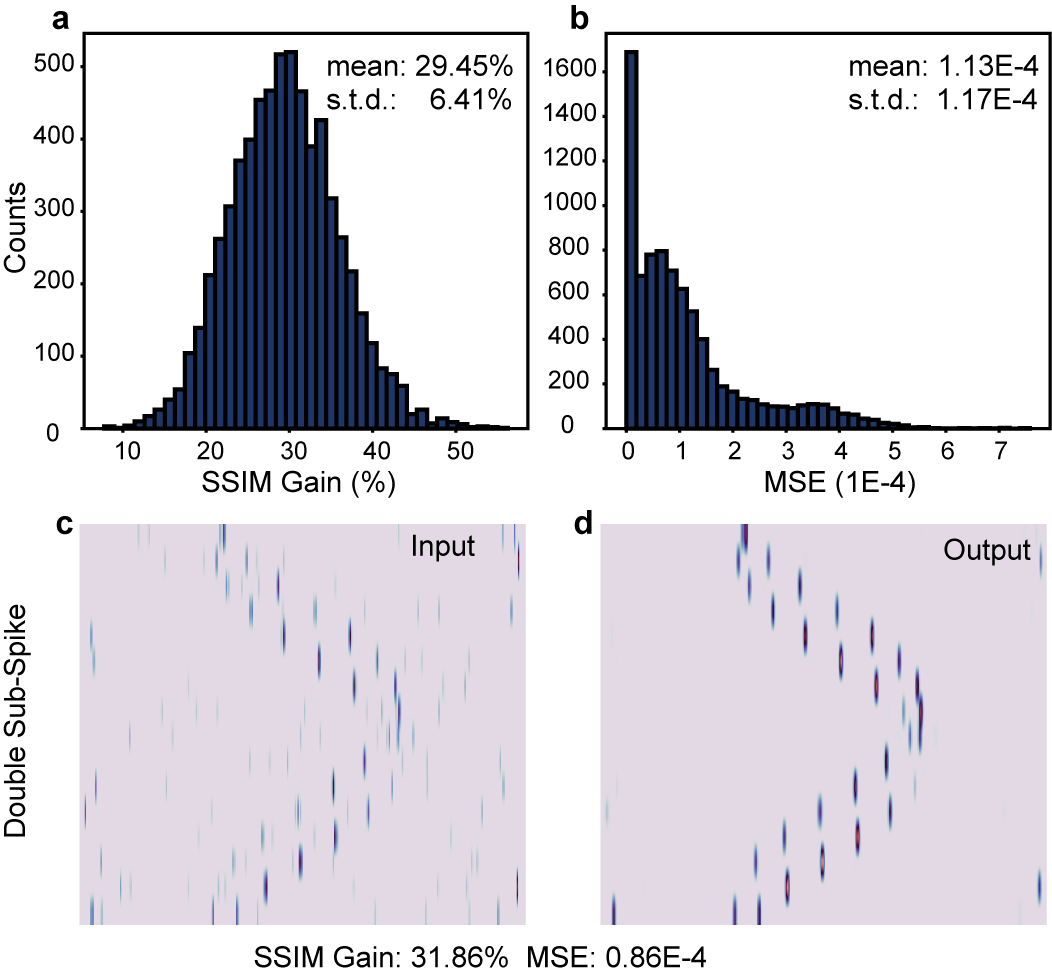}
\caption{\textbf{Denoiser Autoencoder Model Results} a) and b) show error distribution histograms for the SSIM-G metric and MSE, respectively, and c) and d) show the model input and output for a double sub-spike SASE pulse with the respective SSIM-G and MSE values labeled.
}
\label{fig:denoiseAutoModel}
\end{figure}

Fig.~\ref{fig:denoiseAutoModel}a shows the SSIM-G distribution across the test dataset, excluding cases where the ground truth is all zeros, which are explicitly handled by the zero-pulse classifier network (see the Supplementary Material and Fig.~S2).
The SSIM-G distribution is centered around 29\% with all positive values, meaning the denoiser model always improves the sinogram.
For completeness, the mean-squared error (MSE) across the test dataset is shown in Fig.~\ref{fig:denoiseAutoModel}b.
Fig.~\ref{fig:denoiseAutoModel}c and d show a randomly selected double-sub-spike example near the center of the SSIM-G distribution (see Fig.~S3 in the Supplementary Material for more details).

\begin{table*}
\caption{DCIPHR parameters and runtime}
\label{tab:paramsRunTime}
\begin{ruledtabular}
\begin{tabular}{llccc}
\textbf{Model (Identifier)} & \textbf{Component} & \textbf{\# Parameters} & \textbf{Parameter Memory (MB)} & \textbf{Single Inference Runtime (\SI{}{\micro\second})} \\ \hline
Denoiser (1) & Zero Classifier (1a) & 70,345 & 0.28 & 28.2 \\
             & Autoencoder (1b)     & 46,529 & 0.19 & 96.3 \\
Classifier (2) & --                 & 1,458,597 & 5.83 & 61.4 \\
Single-Pulse $\phi$ Regressor (3) & -- & 12,196,240 & 48.78 & 52.2 \\
Double-Pulse $\Delta\phi$ Regressor (4) & -- & 23,330,400 & 93.32 & 72.0 \\ \hline
\textbf{Totals} & -- & \textbf{37,102,111} & \textbf{148.40} & \textbf{168.3} \\
\end{tabular}
\end{ruledtabular}
\end{table*}

Table~\ref{tab:paramsRunTime} summarizes the parameter count and memory footprint for the denoiser's two sub-networks.
In total, the encoder-decoder and zero-pulse classifier require $\sim$117K parameters and occupy 0.47~MB of memory.
The zero-pulse classifier runs in \SI{28.2}{\micro\second} and the encoder-decoder runs in \SI{96.3}{\micro\second} on a Groq card, with a total run time of \SI{96.3}{\micro\second} since the two models can run concurrently.
 
\subsection{Sub-spike Number Classifier}
\begin{figure}[ht!]
\centering\includegraphics[width=.5\textwidth]{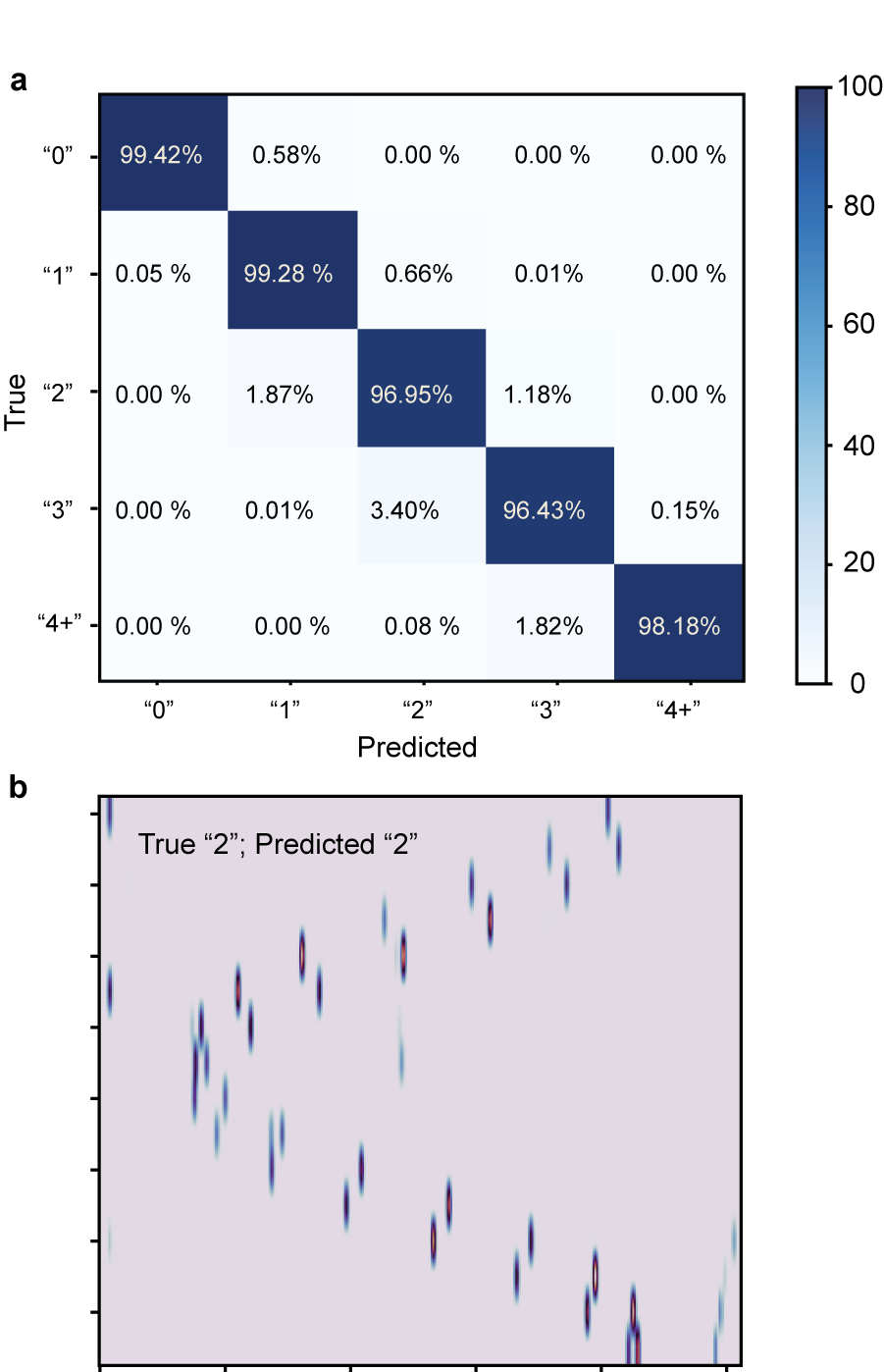}
\caption{\textbf{LSTM Network for Sinogram Featurization Classification}
a) cross entropy matrix for the test set and b) an example of true 2 sub-spike case predicted correctly as 2.}
\label{fig:resultsClass}
\end{figure}

Performance is evaluated on a dataset with an even distribution of sinusoid counts across the 5 classes, 0 to 4+ where the 4+ category can contain between 4 and 10 sinusoids.
We choose an even distribution across classes in spite of the more favorable class distribution of Ref.~\cite{lutman2018high} in order to maintain equally weighted impact on accuracy for all cases expected to be present in an experimental distribution.
Fig.~\ref{fig:resultsClass}a shows the cross-entropy matrix results for true versus predicted.
The model achieves an overall accuracy exceeding 96\%, with notable performance for critical classes: ``0'' and ``1'' sinusoid classes achieve over 99\% accuracy, and the ``2'' sinusoid class exceeds 96\% accuracy. 
Fig.~\ref{fig:resultsClass}b shows a particularly challenging example of the denoised double-sub-spike case correctly classified. 
The Supplementary Material expands on this example, Figs.~S4c--e displaying cases where true ``2'' sub-spikes are classified as ``1,'' ``2,'' or ``3,'' respectively and 
Figs.~S4f--h similarly showing examples where true ``3'' sub-spikes are predicted as ``2,'' ``3,'' or ``4+.''  

Shown in Table~\ref{tab:paramsRunTime}, the classification network comprises $\sim$1.5M parameters, requiring 5.83 MB of memory.  
The single-batch, non-pipelined inference runtime is \SI{61.4}{\micro\second} on a Groq card.

\subsection{Double Sub-spike Regressor}

\begin{figure}
\centering\includegraphics[width=.5\textwidth]{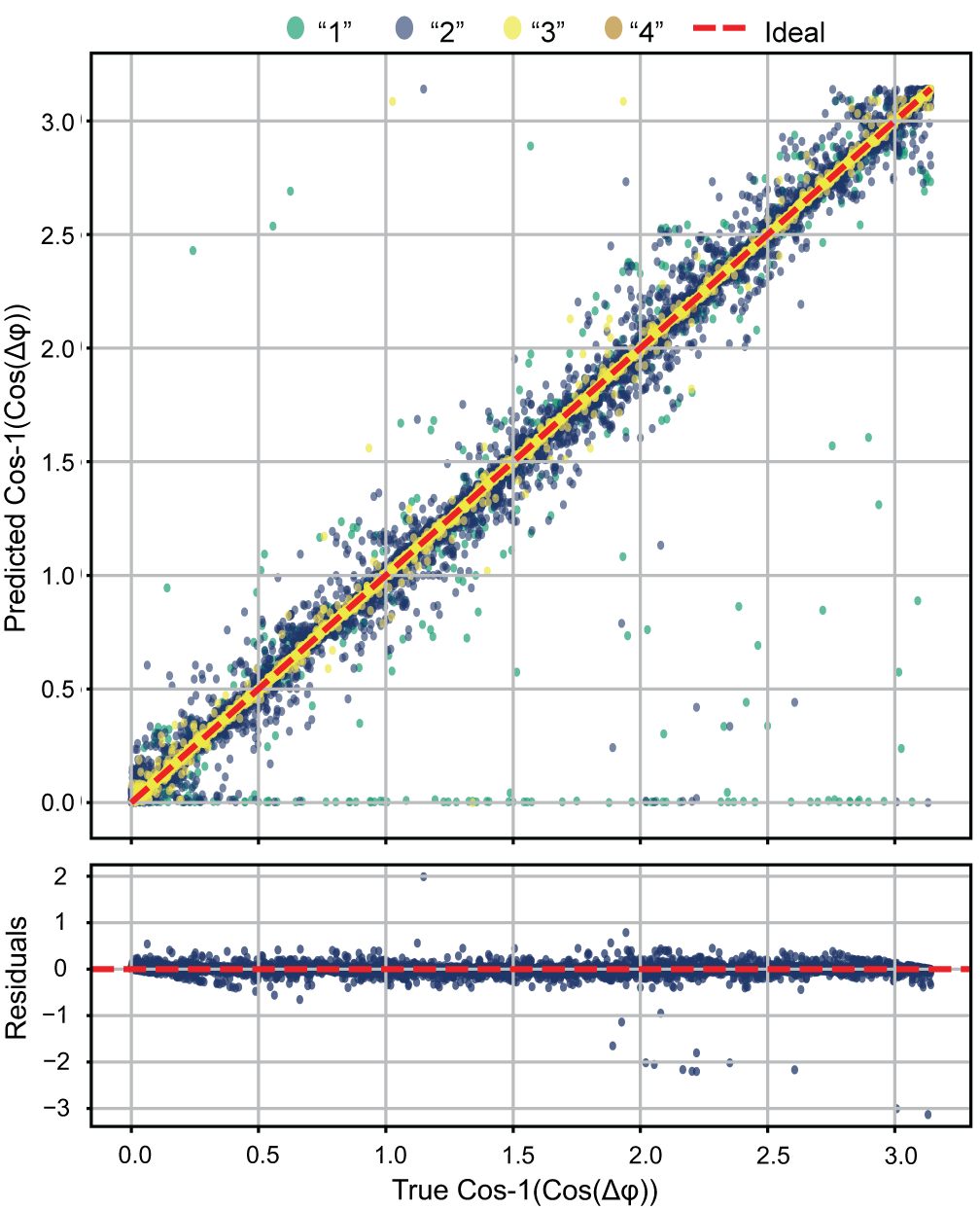}
\caption{\textbf{Double Sub-Spike Phase Difference Regression}
Shows the true versus predicted phase values, the labels from the classifier for each data point, and the residuals for those correctly identified as 2 sub-spikes.}
\label{fig:reg2Results}
\end{figure}

Fig.~\ref{fig:reg2Results} compares predicted and true phase differences, color-coded by Classifier results, along with the associated residuals for the correctly classified two sub-spike events.  
Correctly classified examples exhibit strong agreement between predictions and true values, while misclassifications arise from overlapping or low-intensity sinusoids (see Supplementary Material Fig.~S6c--f).
The final phase difference RMSE for the two-pulse model is 0.04652 radians, calculated using only the correctly classified values.
This phase difference error corresponds to a streaking laser wavelength-dependent temporal resolution as calculated in Eq.~\ref{eq:temporal} where Table~S1 in the Supplementary Material summarizes achievable temporal resolutions for standard infrared laser wavelengths.  
For instance, using a \SI{10.6}{\micro\meter} wavelength yields a resolution of 262 attoseconds, while shorter wavelengths like \SI{1.2}{\micro\meter} achieve a resolution of 29.6 attoseconds, with the trade-off that the total x-ray temporal duration must be shorter. 
The ResNet-34 contains $\sim$23.3 parameters with a memory usage of 93.32~MB and a runtime of \SI{72.0}{\micro\second} per batch on a Groq card.

\section{Discussion}
\label{sec:discussion}
DCIPHR, trained on synthetic data that closely mimics the operational output of the MRCO instrument and the expected typical x-ray laser pulses \cite{lutman2018high}, delivers record temporal resolution for the two sub-spike separation, greater than 96\% accuracy across the 5 classes for sub-spike identification, and signal denoising that guarantees signal quality improvement.
Because DCIPHR is trained on generalized SASE-like sinograms rather than a specific FEL mode, its inference logic is directly applicable to XLEAP’s double-pulse operation, provided that the photon energy and temporal separation satisfy detector constraints.

Additionally, DCIPHR represents several notable innovations.
The bidirectional LSTM scans both vertical directions across image rows to capturing phase and energy variations within sinusoidal patterns. 
This adaptation enhances the network’s capacity to detect and classify sub-spikes with high precision, a critical feature for phase reconstruction in x-ray diagnostics.
Additionally, a critical design decision in the single and double phase regression models within DCIPHR was balancing the trade-off between output class size and discretization granularity of the phase.
For regression tasks, discretization of the continuous phase range into classes permitted precise predictions while minimizing the total learnable parameters.
Generally, classification problems are considered more tractable than regression problems, and mapping the regression task to a classification one enables the use of efficient loss functions while simultaneously reducing the complexity of the problem~\cite{torgo1996regression, salman2012regression}.
For instance, the two sub-spike model’s use of 4000 classes ensured a phase resolution of approximately 0.79 milliradians, exceeding the precision limit of the final network (46.52 milliradians).
This deliberate over-discretization avoided limiting performance due to quantization while maintaining model efficiency.
Moreover, a custom error encoding based on internal angular difference between predicted and true phases improved convergence and phase-wrapping robustness during training. 
For training, the models all used custom schedulers and dataloaders (see Supplementary Material).
Together, these innovations in model design and optimization underscore the meticulous engineering that enabled DCIPHR's unprecedented accuracy.

Optimized for deployment on high-speed AI inference platforms, the DCIPHR model is relatively lightweight yet robust, with 37,102,111 parameters occupying 148.40 MB of memory. 
It delivers high-speed inference, processing each sinogram with a latency of \SI{168.3}{\micro\second} when deployed on Groq cards where the classifier and regression models are run concurrently. 
With simple inter-model pipelining, DCIPHR accomplishes greater than 10~kHz throughput according to Groq's internal deterministic timing analysis; the \SI{96.3}{\micro\second} autoencoder is the bottleneck. 

A throughput that exceeds 10 kHz directly supports the staged increases in repetition rate at LCLS-II.
However, with strategic hardware deployment following the paradigm of parallel and sequentially heterogeneous hardware architectures, the DCIPHR model could be further optimized to the available resources and data pipeline. 
For example, the actual MRCO detector uses a mixture of different hardware data processing components in the data pipeline including FGPA's, Graphic Processing Units (GPU's), and targeted AI inference hardware like the Groq GC1 card.
Different modules of the DCIPHR model are more optimally suited for the specific hardware. 
While the scope of this paper focused on comparing timing all on the same device, any real-time streaming operation would certainly benefit from distribution across architecturally different devices.
For instance, the autoencoder model and zero classifier model within the Denoiser both have relatively high latency compared to their total number of parameters and layers.
The throughput bottleneck is currently the autoencoder model with a latency of \SI{96.3}{\micro\second}.
Since this autoencoder has no skip-connections, both the zero pulse classifier and the encoder portion of the autoencoder, with appropriate quantization, could be transferred to an upstream FPGA optimized for CNNs using the SLAC Neural Network Library (SNL)~\cite{dave2025fpga}. 
With these optimizations, bottleneck latency can be immediately reduced to 72 µs, enabling pipelined throughput approaching 14 kHz.
Moreover, future work could completely drop the decoder portion of the Denoiser and either train downstream modules in DCIPHR on the latent representation or incorporate the decoder layers where needed depending on the task. 
Furthermore, with FPGAs or other targeted hardware, intra-model pipelining can be implemented, given sufficient resources, to dramatically increase the throughput and reduce latency. 
Currently, we are exploring options for distributed processing across FPGAs and Groq cards, quantization for these deployments, further model reduction, and additional pipelining. 
Importantly, scaling to MHz-class throughput does not inherently require next-generation hardware; rather, it depends on optimizing module–hardware matching and enabling intra-model pipelining within existing heterogeneous architectures.

\section{Conclusion}\label{sec13}
The DCIPHR architecture provides a transformative solution for high-throughput, single-shot analysis of images with periodic structure, such as those in x-ray diagnostics. 
It enables parallel real-time data compression, classification, and feature extraction and it significantly reduces storage demands while preserving key characteristics in one- and two-sub-spike events.
It also lays the groundwork for online handling of more complex structures.

These advances are crucial for real-time diagnostics that support autonomous experimental control feedback systems.
For example, in magnetic fusion energy, there is substantial benefit in implementing models at the reactor in order to forecast plasma behavior to ensure stable operation.
At facilities like the Time-resolved Atomic, Molecular, and Optical Science (TMO) endstation at LCLS-II, where MRCO resides and which can currently produce x-ray data at rates around 33 kHz~\cite{walter2022time, pile2024fast}, effective online feature extraction and data compression are critical for real-time autonomous experimental decision support. 
This is necessary to manage the continuous terabits-per-second of data, an increasingly pressing challenge for facilities operating at continuous high rates.

The sub-spike classifier component achieves greater than 96\% accuracy across all classes, for two sub-spike events--pivotal in x-ray pump/x-ray probe studies--DCIPHR achieves a phase separation resolution of 0.04652 radians, translating to an impressive temporal resolution of 29.6 attoseconds when paired with a \SI{1.2}{\micro\meter} streaking laser.
This phase separation performance not only surpasses traditional iterative reconstruction techniques, which suffer from prohibitive latency and throughput constraints, but does so in a fashion compatible with ultra-high-rate streaming data.
With inter-model pipelining among Groq cards, throughput could exceed 10~kHz.
With improved hardware deployment and intra-model pipelining, throughput can be further increased to meet the operational demands which in fact are one to two orders of magnitude higher than achieved here. 

In tandem with these advancements, LCLS-II is pushing the frontier in x-ray repetition rates and energy tunability, emphasizing online x-ray optimization~\cite{edelen2020machine, hanuka2021physics, duris2020bayesian}.
Initiatives are underway to develop comprehensive digital surrogate models of the x-ray generation process, from electron generation at the photoinjector laser to the linear accelerator and x-ray production stages~\cite{hirschman2024design,gupta2021improving}. 
The DCIPHR network’s rapid, in situ processing capabilities align with this larger vision of digital twins and real-time optimization, serving as an integral component for future applications in experimental feedback loops and autonomous facilities.
By bridging ML, hardware optimization, and experimental diagnostics, DCIPHR exemplifies the future of data-driven, real-time control both generally and particularly in high-rate attosecond x-ray science.

% \section{Methods}\label{sec:methods}
% \input{sections/methods}

\section*{Supplementary Material}
The Supplementary Material contains details on the data generation code, additional DCIPHR model results, information on the timing estimates for the AI hardware, and detailed usage instructions for the code repositories. 

\begin{acknowledgments}
The authors acknowledge Greg Stewart, Senior Graphic Designer at SLAC National Accelerator Laboratory, for his contributions to the design and preparation of the main figures and the graphical highlight for this work.
Development and research for MRCO at LCLS is supported by DOE BES under contract No. DE-AC02-76SF00515, DOE BES FWP100498, and for the EdgeML work under DOE BES FWP100643, DOE FES FWP101046 and DOD through an NDSEG Fellowship. 
\end{acknowledgments}

\section*{Data Availability Statement}
The data that support the findings of this study are openly available in the Stanford Digital Repository at \url{https://doi.org/10.25740/bn162bn3804}.
Additionally, new raw data can be generated via the CookieSimSlim code base hosted on GitHub at \url{https://github.com/ryancoffee/CookieSimSlim/tree/release/v1.2}.
The model-specific code is available on GitHub at \url{https://github.com/jhirschm/COOKIE_ML/tree/release/1.0.1}.

\appendix

\nocite{*}
\bibliography{aipsamp}% Produces the bibliography via BibTeX.

\end{document}